# Band-tail shape and transport near the metal-insulator transition in Si-doped $Al_{0.3}Ga_{0.7}As$

Jennifer Misuraca[1,*], Jelena Trbovic[2], Jun Lu[3], Jianhua Zhao[3], Yuzo Ohno[4], Hideo Ohno[4], Peng Xiong[1], Stephan von Molnár[1]

[1] Department of Physics, Florida State University, Tallahassee, Florida 32306, USA

[2] Institute of Physics, University of Basel, Basel, Switzerland

[3] Institute of Semiconductors, Chinese Academy of Sciences, Beijing, China

[4] Research Institute of Electrical Communication, Tohoku University, Sendai, Japan

In the present work, an infrared LED is used to photodope MBE-grown $Si:Al_{0.3}Ga_{0.7}As$, a well known persistent photoconductor, to vary the effective electron concentration of samples *in situ*. Using this technique, we examine the transport properties of two samples containing different nominal doping concentrations of Si ($1 \times 10^{19}$ cm$^{-3}$ for sample 1 (S1) and $9 \times 10^{17}$ cm$^{-3}$ for sample 2 (S2)) and vary the effective electron density between $10^{14}$ and $10^{18}$ cm$^{-3}$. The metal-insulator transition (MIT) for S1 is found to occur at a critical carrier concentration of $5.7 \times 10^{16}$ cm$^{-3}$ at 350 mK. The mobilities in both samples are found to be limited by ionized impurity scattering in the temperature range probed, and are adequately described by the Brooks-Herring screening theory for higher carrier densities. The shape of the band-tail of the density of states (DOS) in $Al_{0.3}Ga_{0.7}As$ is found electrically through transport measurements. It is determined to have a power law dependence, with an exponent of -1.25 for S1 and -1.38 for S2.

[*] correspondence should be addressed to jm05h@fsu.edu

# 1. MOTIVATION AND INTRODUCTION

In the past decade, the demonstration of optical spin injection in GaAs established that the spin lifetime in that material is significantly enhanced around the metal to insulator phase transition (MIT).[1] Comparable lifetimes have also been observed through electrical spin injection and detection in GaAs doped with densities near the critical carrier concentration.[2,3] The sensitivity to doping of spin lifetime in GaAs is difficult to investigate in great detail, however, because replicas need to be prepared in order to study differently doped samples. This problem can be circumvented in certain alloy composition ranges of $Al_xGa_{1-x}As$ where persistent photoconductivity (PPC) occurs.[4] Utilizing PPC, one can perform measurements at various carrier concentrations spanning orders of magnitude on the *same* sample without the inaccuracies associated with using replicas. The ultimate goal of investigating this material is to conduct a systematic, comprehensive study of spin diffusion and its dependence on carrier concentration. Since AlGaAs is structurally very similar to GaAs,[5] research on doping densities in AlGaAs should help to optimize spin injection experiments that use GaAs channels. In this paper, the material which will be used for this long-term objective is characterized using PPC. We focus on the transport properties of the bulk AlGaAs material and how these change with carrier density. In view of the fact that spin diffusion length is longest near the MIT, this is the region where the investigation is focused.

PPC is a phenomenon in which materials experience an enhancement in conductivity when a photoexcitation occurs which remains long after the excitation is discontinued. It is known to occur in an assortment of III-V and II-VI semiconductors, such as AlGaAs and GaAsP,[6] CdMnTe and CdZnTe,[7] and ZnCdSe,[8] when each is doped with wide variety of donors. A detailed review of PPC in III-V semiconductors was presented by Mooney.[9]

This paper begins with a brief presentation of the experimental methods and transport results, including a simplified explanation of PPC and how the phenomenon is used to effectively increase the carrier concentration in $Al_{0.3}Ga_{0.7}As$. The PPC-driven MIT in $Al_{0.3}Ga_{0.7}As$ is presented next, as the remainder of the research will be focused on the insulating side of this transition. After that, the scattering mechanisms are discussed and analyzed. Finally, the shape of the band-tail of the density of states (DOS) in the material is determined. In general, the exact nature of the band-tail in highly disordered systems is notoriously hard to elucidate experimentally. However, by using PPC and analyzing the transport data on the insulating side of the MIT, the shape of the band-tail of the DOS in this material emerges.[7]

## 2. EXPERIMENTAL METHODS AND RESULTS

The structure of the MBE-grown Si:Al$_{0.3}$Ga$_{0.7}$As wafer is shown in Fig. 1(a), where the active layer has a nominal Si-doping of 1×10$^{19}$ cm$^{-3}$ in sample 1 (S1) and 9×10$^{17}$ cm$^{-3}$ in sample 2 (S2). Hall devices with dimensions L = 200 μm and W = 75 μm are patterned using standard photolithographic and wet chemical etching techniques; an optical image of the Hall bar is shown in Fig. 1(b). The 20 nm thick cap layer of highly-doped GaAs ensures good electrical connection between the Al$_{0.3}$Ga$_{0.7}$As and the AuGe contacts and is subsequently etched away after the contacts are annealed. The wired samples are then mounted onto a cryostat sample holder along with commercial infrared (IR) LEDs (λ=940 nm) which are positioned to directly illuminate the active area of the Hall bars. Samples are measured in a Janis $^4$He cryostat for temperatures T > 5 K and MIT studies are done in an Oxford $^3$He cryostat down to 350 mK.

In Si:Al$_{0.3}$Ga$_{0.7}$As, PPC is attributed to the donor having two metastable states, a shallow hydrogenic state and a deep localized state (referred to as a DX center)[4, 10-12] which can be represented by the configuration coordinate (CC) diagram shown in Fig. 2(a).[7, 10] When the sample is cooled in the dark, the electrons are captured by the DX centers of the Si. Since the electrons in these highly localized deep states do not contribute to the conduction, the material becomes very resistive at low temperatures (see the top curve in Fig. 2(b)). At 5 K, illumination by the IR LED creates a photoexcitation that will cause the DX centers to transfer charges into the shallow states. Infrared light is used because its energy ($E_{IR}$= 1.32 eV) exceeds the optical energy required, $\Delta E_o$, for the Al$_{0.3}$Ga$_{0.7}$As samples but is smaller than the band gap ($E_g$=1.80 eV at 300 K)[13] to prevent excitations from the valence band. Although $\Delta E_o$ is not easily determined for the material, it is known to be larger than the thermal electron emission energy ($E_e$= 0.43 eV)[4] and smaller than $E_g$. Once the electrons are excited into the shallow states, the number of mobile carriers in the system is increased as is the effective carrier concentration of the material;[7, 14] this means that the time in which the LED is turned on is correlated with the effective carrier concentration of the material.

Once the LED is turned off, the material remains conductive below ~80K, which is where the thermal fluctuations begin to overcome the recapture barrier $\Delta E_R$ (Fig. 2(a)). DC and AC lock-in techniques are used to determine longitudinal resistance, $R$, and Hall resistance, $R_H$, (see Fig. 1(b)) of the devices as a function of temperature for each illumination time from 5 K to 165 K. At this higher temperature, the PPC effect completely vanishes because the DX centers recapture the electrons again, and the illuminated sample begins to mimic the dark sample (bottom curve in Fig. 2(b)). This means that the entire process is reversible and can be repeated for many different illumination times by simply heating the sample above this critical temperature and cooling down again in the dark.[7, 14]

The conductivity (approximated as $\sigma = 1/\rho$) versus inverse temperature is shown on a log scale in Fig. 3 in the insulating regime and near the MIT. This approximation is valid since the off-diagonal terms in the resistivity tensor are small compared to the diagonal ones. The conductivities closer to saturation are many orders of magnitude larger than the curves shown. The three dimensional carrier concentration, $n = \dfrac{-IB}{V_H e t}$, is plotted on a log scale versus inverse temperature in Fig. 4, where $V_H$ is the Hall voltage, I is the current, B is the perpendicular magnetic field, $e$ is the electronic charge, and $t$ is the thickness of the sample.[15] In all measurements, $V_H$ was linear in $B$. The curves depicted in Fig. 4 demonstrate that between these two highly Si-doped $Al_{0.3}Ga_{0.7}As$ samples, up to four orders of magnitude in carrier concentration may be accessed by photodoping. S1 has a dark carrier concentration on the order of $10^{16}$ cm$^{-3}$ and S2 has a much lower dark carrier concentration on the order of $10^{14}$ cm$^{-3}$. From these transport data, important information about the critical carrier concentration of the MIT, the scattering mechanisms which suppress the mobility of the electrons, and the shape of the band-tail of the DOS in this highly disordered material may be extracted.[7]

## 3.  DISCUSSION

### A. Insulator to metal transition

The resistivity of S1 was measured versus temperature from 10 K to 350 mK on the insulating side and on the conducting side near the transition to determine the critical carrier concentration for the MIT. The conductivity of Si-doped AlGaAs is shown experimentally[14] to vary at low temperatures according to:

$$\sigma(T,n) = \sigma(0,n) + m(n)T^{1/2} \qquad (1)$$

where $m$ is a coefficient that does not depend on temperature, $n$ is the effective carrier concentration, and the $T^{1/2}$ term comes from electron-electron interactions. In S1, we see this dependence emerge below 2.25K (see Fig. 5(a)). It is clear from Fig. 5(a) that the $T^{1/2}$ dependence is well obeyed in this material and temperature range all the way up to the MIT. Similar observations have been made by Katsumoto et al.[14] in AlGaAs (doped Si:$1\times10^{18}$ cm$^{-3}$) and Bishop et al.[16] in NbSi, and in both cases this was utilized to identify the point of the MIT.

According to the scaling theory of localization[17,18] the MIT is continuous and scales as

$$\sigma(T=0,n) = \sigma_0 (n/n_c - 1)^\nu \qquad (2)$$

where $\sigma(T=0, n)$ is the zero temperature conductivity, $\sigma_0$ is a constant, $n_c$ is the critical carrier concentration, and $\nu$ is the critical conductivity exponent which is 1 for AlGaAs.[14] Thus, at the MIT (that is, when $n = n_c$), the zero temperature conductivity becomes zero. In Fig. 5(a), the conductivity of S1 is plotted against $T^{1/2}$ for curves in close proximity to the transition and the zero temperature conductivity is extrapolated from a linear fit for each illumination time (dashed line). In Fig. 5(b), these zero temperature conductivities are plotted versus illumination time and the MIT is determined to occur at 300 s of illumination (at 5K using 50μA of current through the LED). Since the MIT is defined only at zero temperature, the Hall measurement was taken at the lowest temperature that can be achieved in the cryostat. The $n_c$ was determined to be $5.7\times10^{16}$ cm$^{-3}$ at 350 mK, which is below the upper limit of $n_c$ obtained by using the Mott criterion for Al$_{0.3}$Ga$_{0.7}$As ($n_c < 1.33\times10^{17}$cm$^{-3}$).[13,19] However, this value is not consistent with the previous work by Katsumoto et al.[14] whose Si doping level is an order of magnitude lower ($1\times10^{18}$ cm$^{-3}$), but is otherwise equivalent in structure. In this instance, a critical carrier concentration of $2.6 \times 10^{16}$ cm$^{-3}$ was determined via resistivity versus temperature measurements taken from 350 mK to 40 mK using Hall data that was obtained at 77 K.[14] In a subsequent investigation on the same material[20], when Hall data was measured at lower temperatures (see Fig. 6 in Ref. 20), a higher $n_c$ is indicated but is still inconsistent with our results.

The carrier concentration in S1 is shown to vary strongly even at these very low temperatures, and shows an increase from $5.7 \times 10^{16}$ cm$^{-3}$ at 350 mK to $9.0 \times 10^{16}$ cm$^{-3}$ at 5 K. The reason for this increase is not clear, although a similarly, but less pronounced increase is also seen in the detailed Hall measurement versus temperature study by Katsumoto from 100 mK to 300 mK.[20]

### B. Temperature dependence of the mobility

The mobility ($\mu = \sigma/ne$) for each sample is calculated from the transport data in Figs. 3 and 4 and found to be dominated by ionized impurity scattering in the temperature range measured (30 K to 110 K).[21] The mobility is shown for both samples in Fig. 6 in the temperature range from 50 K - 100 K for clarity. The mobility limited by ionized impurity scattering $\mu_{II}$[22] has a characteristic temperature dependence given by:

$$\mu_{II} = \frac{C_{II}}{N_{II}} T^{3/2} \qquad (3)$$

where $N_{II}$ is the ionized donor density and $C_{II} = \frac{2^{7/2} \varepsilon^2 k_B^{3/2}}{\pi^{3/2} e^3 m_{eff}^{1/2}} \frac{1}{\ln\{[1 + 9(\varepsilon d_I k_B T)^2]/e^4\}}$, where $k_B$ is the Boltzmann constant and $d_I$ is the average distance between the neighboring scattering centers.[22] Using $\varepsilon = 12.2\varepsilon_0$ for the dielectric constant and $m_{eff} = 0.09 m_0$ for the effective mass,[13] $C_{II} = 6 \times 10^{15}$ (m$^2$/Vs)m$^{-3}$K$^{-3/2}$ for AlGaAs. Even though there is a temperature dependence in $C_{II}$ as shown, it is very weak and can be considered negligible.

Using this model (Eq. 3), we find agreement with the temperature dependence of the mobility of S2 for the dark curve (blue line in Fig. 6). The temperature exponent is 3/2 and the ionized donor density $N_{II}$ is a free fitting parameter with a value of $1.1 \times 10^{14}$ cm$^{-3}$. This is in good agreement with the number of free carriers as determined by Hall effect measurements in this sample with no illumination. However, curves with higher photodoping in the same sample (S2) begin to deviate from the $T^{3/2}$ dependence and cannot be modeled in this way.

To account for this behavior at higher carrier densities, the Brooks-Herring (BH) screening model is applied. The BH formula[23] for carrier mobility when the screening of the ionized impurities is taken into account is given by:

$$\mu_{BH} = \frac{C_1}{N_{II}} T^{3/2} \left[ \ln\left( C_2 \frac{T^2}{n} \right) \right]^{-1} \quad (4)$$

where $n$ is the uniform electron concentration, $C_1 = \frac{128\sqrt{2\pi}\varepsilon^2 k_B^{3/2}}{\sqrt{m_{eff}} Z^2 e^3}$ and $C_2 = \frac{24 m_{eff} \varepsilon k_B^2}{e^2 \hbar^2}$ are constants, and $Z$ is the impurity charge.[23] For AlGaAs, $C_1 = 1 \times 10^{19}$ (m$^2$/Vs)m$^{-3}$K$^{-3/2}$ and $C_2 = 1.5 \times 10^{20}$ m$^{-3}$K$^{-2}$. Using the BH formula, higher photodoped curves for S2 can be successfully fit, yielding higher ionized donor densities, in good agreement with the free carrier densities determined for these curves.

According to the ionized impurity scattering model (Eq. 3), the mobilities in S1 should be lower than S2 by a factor of 10; however only a decrease of about a factor of two is observed. Thus the ionized impurity screening model cannot be used to fit even the dark curve for S1. The BH model (Eq. 4), however, can be fit to yield a $N_{II}$ of $5 \times 10^{16}$ cm$^{-3}$, which is on the order of the free carrier density obtained via Hall measurements for this curve (red line in Fig. 6). The BH model continues to be valid for even higher photodoped curves until $N_{II}$ increases to approximately $9 \times 10^{16}$ cm$^{-3}$. Further increasing of the photodoping (and thus lowering of the effective temperature exponent) cannot be explained within the BH model.

### C. Density of states

The determination of the shape of the band-tail of the DOS in Al$_{0.3}$Ga$_{0.7}$As on the insulating side of the transition follows the approach of Terry et al.[7] Both samples are used in order to span as large an energy range as possible. The carrier concentration $n$ is given by:

$$n = n_\infty e^{\frac{-E_s}{k_B T}} \quad (5)$$

where $n_\infty$ is the infinite temperature carrier concentration and $E_s$ is the Hall activation energy, which is the energy required to excite an electron into an extended state. $E_s = E_\mu - E_F$, where the mobility edge, $E_\mu$, is defined as the level at zero temperature for which all electron states at energies smaller than it are localized and all states larger than it are extended; $E_F$ is the Fermi energy of the material. The carrier concentration data in Fig. 4 is presented on a log scale versus inverse temperature in order to easily extract the infinite temperature carrier concentrations and the Hall activation energies in the samples (from Eq. 5). In the dark, the Fermi energy is low and inside the band-tail. As the sample is illuminated, the Fermi energy increases systematically toward the mobility edge, and the MIT occurs when the two energies are equal. This means that the Fermi energy may be tuned systematically by photodoping the sample, thus decreasing the Hall activation energies with illumination.

The high temperature data of Fig. 4 allows one to extract information about the deep levels in the material. The linear fit of the slope for each sample (dashed line in Fig. 4) can be used to determine the activation energy of the deep level, $E_{DD}$. The values obtained are 69 meV for S1 and 89 meV for S2. This means that in the lower doped samples, the DX centers are further separated from the mobility edge than in the higher doped samples. These values agree well with deep state energies previously published in the literature for similarly doped AlGaAs samples.[24, 25]

The linear fit of the lower temperature data (solid line in Fig. 4) allows one to extract the $n_\infty$ from the intercept at $1/T = 0$ and the $E_s$ of the shallow states from the slope of the line for each illumination time. The Hall activation energy at low temperature in the dark indicates some population of the shallow levels with values of 4 meV for S1 and 15 meV for S2. This means that the shallow states of the higher doped sample are closer to the mobility edge than those of the lower doped sample. These values also agree well with previously reported data from Chand et al. for similar AlGaAs samples.[24] In order to determine the shape of the DOS, these shallow Hall activation energies and corresponding infinite temperature carrier concentrations must be determined for many different illumination times. Using this method, the maximum amount of the band-tail that we can probe is from near the mobility edge, $E_s = 0$, to the Fermi energy of the unilluminated, lowest-doped sample, $E_s = 15$ meV.

Establishing the shape of the DOS requires that a relationship be found between the Hall activation energies and the infinite temperature carrier concentrations, since the DOS is the derivative of the carrier concentration with respect to $E_F$. A reasonable assumption is a power law, i.e. $n_\infty = CE_s^\beta$, where $C$ and $\beta$ are constants. If the plot of $\log(n_\infty)$ versus $\log(E_s)$ is linear, the slope will be equal to the exponent $\beta$ and the constant $C$ can be determined from the intercept. The plot of $\log(n_\infty)$ versus $\log(E_s)$ for S1 is shown in Fig. 7. This provides credence to the assumed power law dependence. The DOS can be found for each sample by taking the derivative of $n_\infty$ with respect to the Fermi energy:

$$N(E) = \frac{dn_\infty}{dE_F} = -C\beta E_s^{\beta-1} \tag{6}$$

The results are:

$$N_1(E) = 1.1 \times 10^{15} \, cm^{-3} eV^{0.25} E_s^{-1.25} \tag{7}$$

$$N_2(E) = 2.6 \times 10^{14} \, cm^{-3} eV^{0.38} E_s^{-1.38}$$

for S1 and S2 respectively. This functional form of the DOS is valid only in the energy range of the experimental investigation (see inset of Fig. 8). The DOS is plotted in Fig. 8 for the experimental Hall activation energies for each sample, and one can see that 15 meV of the band-tail has been investigated. The exponent changes by approximately 10% between samples, creating a steeper curve lower in the band-tail. From Fig. 8, one can see that S1 has a larger DOS than S2 (see e. g. the values at $E_s \sim 4$ meV); this implies that we might expect a larger $n_c$ for the MIT in a higher doped sample than in a lower doped one. This is consistent with the $n_c$ obtained for S1 in Sec. 3A., which is larger than in the previously mentioned work on a lower doped sample.[14]

## 4. CONCLUSIONS

Due to the metastable nature of the Si donor in $Al_{0.3}Ga_{0.7}As$, an infrared LED can be used to populate shallow states and increase the effective carrier concentration of this material *in situ*. This allows measurements to be taken at various effective carrier densities *on the same sample* without removing it from the cryostat, and thus it will greatly reduce the uncertainty that is involved with using replicas. Persistent photoconductive effects in AlGaAs will allow for the determination of the optimal carrier concentration for spin injection experiments in AlGaAs channels, and by analogy in GaAs. This paper has determined how the transport properties of the bulk material are altered in a range of varying carrier densities which will be important for the spin injection and detection experiment. Also, this method of tuning the carrier concentration permits a unique method of electrically determining the band-tail shape of the DOS.

The MIT was explored for the highly doped sample; it was found to occur at a critical carrier concentration of $n_c$ = 5.7x$10^{16}$ cm$^{-3}$ at 350 mK, the lowest available experimental temperature. It is found that ionized impurity scattering and subsequent screening of ionized impurities limits the mobilities of both samples when free carriers are added. The shape of the band-tail of the DOS in this highly disordered semiconductor has a power law dependence, with an exponent of -1.25 in S1 and -1.38 in S2. The magnitude of the exponent increases as the energies get further away from the mobility edge. We showed that $Al_{0.3}Ga_{0.7}As$ can be tuned through the MIT using PPC and can thus offer an ideal spin transport medium for an *in situ* study of spin relaxation versus carrier density near the MIT.

## ACKNOWLEDGEMENTS

We would like to thank Dr. Vladimir Dobrosavljevic for productive discussions about insulator to metal transitions. We are grateful to Gian Salis and D.D. Awschalom for providing early optical spin lifetime characterization of AlGaAs. This work has been supported by NSF grant DMR-0908625 and NSFC 10920101071. J. Trbovic is also grateful to the Swiss NSF for financial support during the preparation of the manuscript.

## REFERENCES

[1] J.M. Kikkawa and D. D. Awschalom, Phys. Rev. Lett. **80**, 4313 (1998).
[2] X. Lou, C. Adelmann, S. Crooker, E. Garlid, J. Zhang, K. S. M. Reddy, S. D. Flexner, C. Palstrom, P. Crowell, Nature Physics **3**, 197 (2007).
[3] C. Awo-Affouda, O. M. J. van 't Erve, G. Kioseoglou, A. T. Hanbicki, M. Holub, C. H. Li, and B. T. Jonker, Appl. Phys. Lett. **94**, 102511 (2009).
[4] P. M. Mooney, T. N. Theis, and E. Calleja, J. of Electron. Mater. **20**, 23 (1991).
[5] Optically excited spins in AlGaAs do not show degradation of spin lifetime compared to GaAs (private communication – Gian Salis and D.D. Awschalom)

[6] D. V. Lang. *Deep Centers in Semiconductors*, 2$^{nd}$ Ed. ed. S. Pantelides. (NewYork: Gordon and Breach, 1992) pp. 591-641.

[7] I. Terry, T. Penney, S. von Molnár, and J. M. Rigotty, Solid State Commun. **84**, 235 (1992).

[8] H. X. Jiang, G. Brown, and J. Y. Lin, J. Appl. Phys. **69**, 9 (1991).

[9] P. M. Mooney, J. Appl. Phys. **67**, R1 (1990).

[10] D. Redfield and R. Bube. *Photoinduced Defects in Semiconductors*. (Cambridge: Cambridge University Press, 1996).

[11] T. N. Theis, P. M. Mooney, and B. D. Parker, J. of Electron. Mater. **20**, 35 (1991).

[12] D. V. Lang, R. A. Logan, and M. Jaros, Phys. Rev. B 19, 1015 (1979).

[13] S. Adachi, J. Appl. Phys. **58**, R1 (1985).

[14] S. Katsumoto, F. Komori, N. Sano, and S. Kobayashi, J. Phys. Soc. Jpn. **56**, 2259 (1987).

[15] R. S. Popovic. *Hall Effect Devices*. (Bristol: Institute of Physics Publishing, 2004) p. 69.

[16] G. Hertel, D. J. Bishop, E. G. Spencer, J. M. Rowell, and R. C. Dynes, Phys. Rev. Lett. **50**, 743 (1983).

[17] E. Abrahams, P.W. Anderson, D.C Licciardello, T.V. Ramakrishnan, Phys. Rev. Lett. **42**, 673 (1979).

[18] C. Leighton, I. Terry, and P. Becla, Europhys. Lett. **42**, 67 (1998).

[19] C. Kittel. *Introduction to Solid State Physics*. (New York: John Wiley and Sons, 2005).

[20] S. Katsumoto, "Photo-induced metal-insulator transition in AlGaAs". Inst. Phys. Conf. Ser. No 108. Paper presented at the Localisation 1990 Conference held at Imperial College, London 13-15 August 1990.

[21] S. Adachi. *GaAs and Related Materials*. (Singapore: World Scientific, 1994).

[22] H. F. Wolf. *Semiconductors*. (New York: John-Wiley & Sons, 1971).

[23] D. Chattopadhyay and H.J. Queisser, Rev. Mod. Phys. **53,** 745 (1981).

[24] N. Chand, T. Henderson, J. Klem, W. T. Masselink, R. Fischer, Y.C. Chang, and H. Morkoc, Phys. Rev. B. **30**, 4481 (1984).

[25] H. Künzel, K. Ploog, K. Wünstel, and B. L. Zhou, J. of Electron. Mater. **13**, 281 (1984).

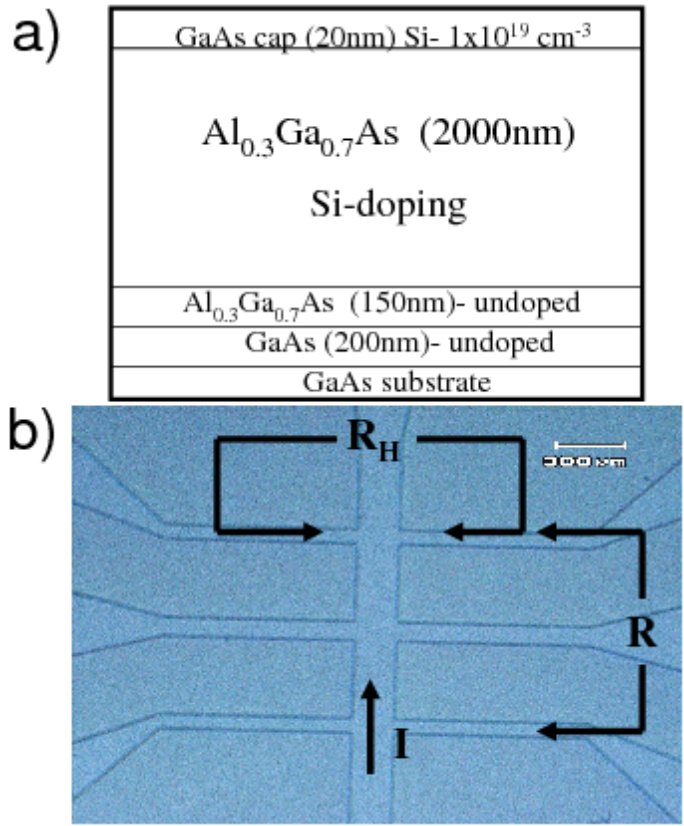

FIG. 1. (a) Structure of the $Al_{0.3}Ga_{0.7}As$ wafers used; the active layer has Si-doping of $1\times10^{19}$ cm$^{-3}$ and $9\times10^{17}$ cm$^{-3}$ for S1 and S2 respectively. (b) Optical image of Hall bar showing where resistance $R$ and Hall resistance $R_H$ are measured.

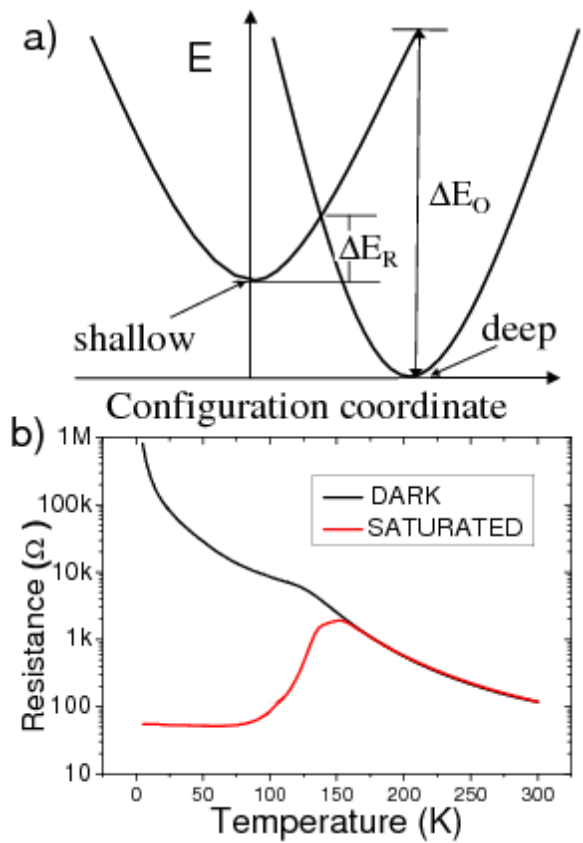

FIG. 2. (a) Configuration coordinate diagram for the metastability of Si in $Al_{0.3}Ga_{0.7}As$; $\Delta E_o$ is the optical energy required to excite charges to the shallow states and $\Delta E_R$ is the energy barrier for recapture by the deep state.[7,10] (b) $R$ vs. $T$ for $Al_{0.3}Ga_{0.7}As$ Hall bar plotted on a log scale for clarity; sample cooled down in the dark (top curve) and saturated sample warmed up (bottom curve).

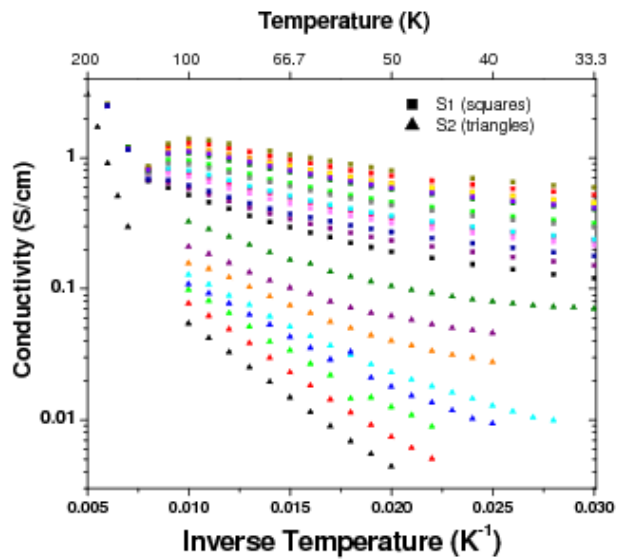

FIG. 3. (color online) Conductivity of S1 (squares) and S2 (triangles) versus inverse temperature for different illumination times from 30 K to 200 K, plotted on a log scale for clarity.

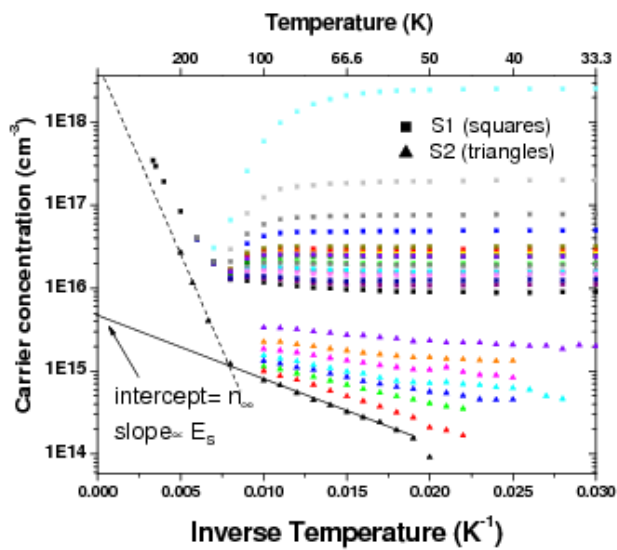

FIG. 4. (color online) Carrier concentration of S1 (squares) and S2 (triangles) versus inverse temperature for different illumination times plotted on a log scale. A linear fit of the lower temperature data allows for the extraction of the infinite temperature carrier concentrations $n_\infty$ and the shallow Hall activation energies $E_s$ for each illumination time (solid line). The linear fit of the high temperature data corresponds to the activation energy of the deep donor, $E_{DD}$, for S2 (dashed line).

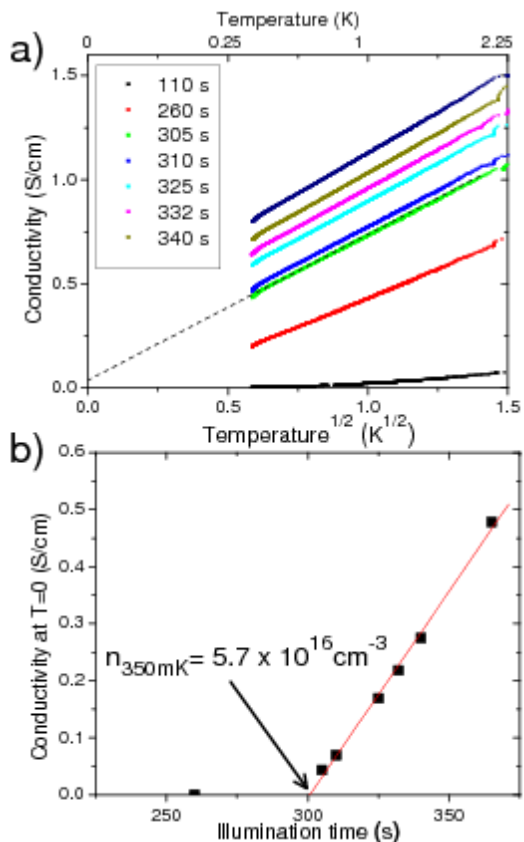

FIG. 5. (color online) (a) Conductivity plotted as a function of $T^{1/2}$ for S1 in close proximity to the MIT below 2.25 K. (b) The zero temperature extrapolations of the conductivity versus illumination time on the metallic side of the transition.

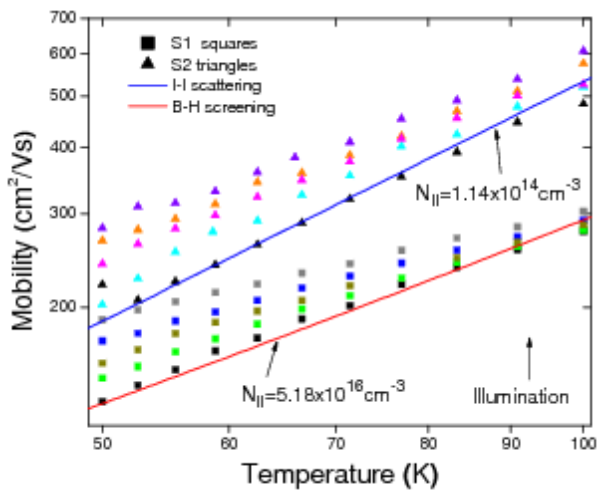

FIG. 6. (color online) Log-log plot of the temperature dependence of the mobility for S1 and S2. The dark S2 curve (black triangles) is fitted with the ionized impurity (I-I) screening model. The dark S1 curve (black squares) is fitted with the Brooks-Herring screening model which includes ionized impurities.

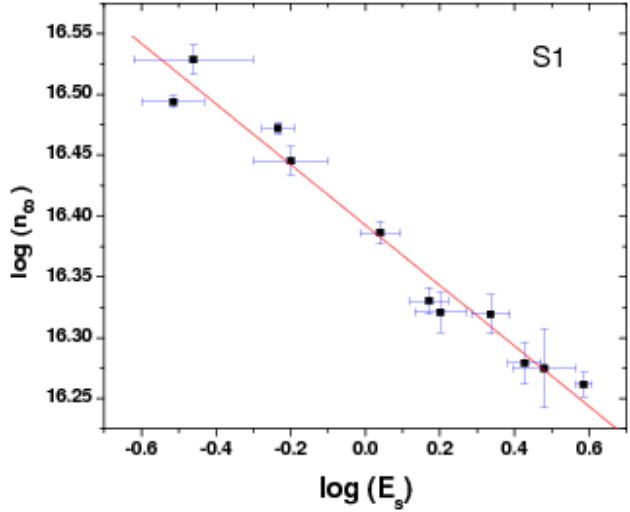

FIG. 7. Log-log plot of infinite temperature carrier concentration versus Hall activation energy for S1, with a linear fit to extract the exponent β and constant C.

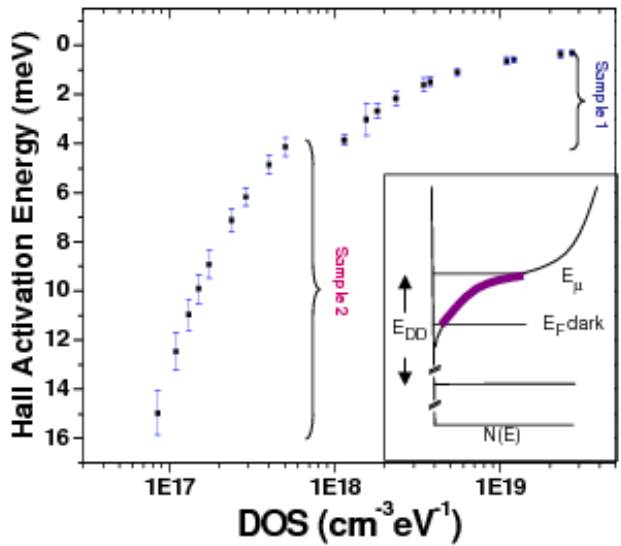

FIG. 8. The energy dependence of the band-tail of the DOS in $Al_{0.3}Ga_{0.7}As$ for S1 and S2. Inset: DOS of a highly disordered semiconductor, where $E_\mu$ is the mobility edge, $E_F$ is the Fermi energy, and $E_{DD}$ is the deep donor activation energy. The portion of the band-tail investigated is shown in bold.